\documentclass[aps,pra,preprint]{revtex4}
\usepackage{graphicx}
\usepackage{amsfonts}
\usepackage{amssymb}
\usepackage{amsmath}

\begin{document}

\title{\bf Soliton oscillations in collisionally inhomogeneous attractive Bose-Einstein condensates \\
}

\author{
P.\ Niarchou$^{1}$,
G.\ Theocharis$^{1}$, 
P.G.\ Kevrekidis$^{2}$, 
P.\ Schmelcher$^{3,4}$, and 
D.J.\ Frantzeskakis$^{1}$ 
}
\affiliation{
$^{1}$ Department of Physics, University of Athens, Panepistimiopolis,Zografos, Athens 157 84, Greece \\
$^{2}$ Department of Mathematics and Statistics,University of Massachusetts, Amherst MA 01003-4515, USA \\
$^{3}$ Theoretische Chemie, Institut f\"ur Physikalische Chemie, 
Universit\"at Heidelberg, INF 229, 69120 Heidelberg, Germany \\
$^{4}$ Physikalisches Institut, Philosophenweg 12, Universit\"at Heidelberg, 69120 Heidelberg, Germany 
}

\begin{abstract}
We investigate bright matter-wave solitons in the presence of a spatially varying nonlinearity. 
It is demonstrated that a translation mode is excited due to the spatial inhomogeneity and its 
frequency is derived analytically and also studied numerically. Both cases of purely one-dimensional and ``cigar-shaped'' 
condensates are studied by means of different mean-field models, and the oscillation frequencies 
of the pertinent solitons are found and compared with the results obtained by the linear stability analysis.
Numerical results are shown to be in very good agreement with the corresponding analytical predictions.

\end{abstract}

\maketitle

\section{Introduction and model}

In a mean-field theoretical framework the macroscopic wavefunction of a 
dilute gaseous Bose-Einstein condensate (BEC) is governed by a classical nonlinear evolution 
equation, namely the Gross-Pitaevskii (GP) equation \cite{dalfovo}. The nonlinearity 
in the GP model is introduced by the interatomic interactions, which are taken into regard through 
an effective mean field. The coefficient (coupling constant) $g$ of the nonlinear term 
in the GP equation is controlled by the $s$-wave scattering length $a$, whose sign and magnitude 
determine many of the fundamental properties of Bose-Einstein condensates (BECs) such as their shape, 
collective excitations or statistical fluctuations \cite{dalfovo}. Importantly, the 
scattering length can be tuned by an external magnetic \cite{Weiner03}, optical \cite{Theis04}, 
or dc-electric \cite{electric} field. The possibility of controlling the interatomic interactions 
in BECs has inspired many experimental and theoretical studies. The former include 
(among others) the generation of bright matter-wave solitons \cite{brightexp1,cew}, 
while the latter predict that time-dependent scattering lengths can be employed to arrest 
collapse in two \cite{FRM1} and three \cite{FRMr} dimensional attractive BECs or to create robust 
solitons \cite{FRM2}, periodic waves \cite{FRMp} and shock waves \cite{FRMs}.

On the other hand, recently, the possibility of varying the scattering length {\it spatially}  
has been proposed, utilizing, e.g., an inhomogeneous external magnetic field in 
the vicinity of a Feshbach resonance \cite{g1}. These, so-called, ``collisionally inhomogeneous'' BECs  
have recently attracted much attention, as they are relevant to many interesting  
applications such as adiabatic compression of matter-waves \cite{g1,fka}, Bloch oscillations of 
matter-wave solitons \cite{g1}, atomic soliton emission \cite{v1}, enhancement of transmittivity 
of matter-waves through barriers \cite{g2,fka2}, dynamical trapping of matter-wave solitons \cite{g2}, 
and so on. Moreover, rigorous mathematical results concerning the existence and the stability of solutions 
of the relevant GP equations also appeared \cite{wein}. 

In this work, we consider an elongated attractive BEC, with inhomogeneous interatomic interactions, which 
is confined in a highly anisotropic harmonic trap, such that the condensate is confined solely in the 
transverse ($r = \sqrt{x^{2}+y^{2}}$) direction and free in the longitudinal ($z$) one. 
Then, its longitudinal mean-field wavefunction $\phi(z,t)$  
satisfies the following normalized one-dimensional (1D) GP equation \cite{g1,g2},  
\begin{equation}  
i \partial_{t} \phi(z,t) =\left[-\frac{1}{2} \partial_{z}^{2} - g(z)|\phi(z,t)|^2\right]\phi(z,t), 
\label{1dGPE}
\end{equation}
where space, time and density are respectively measured in units of the transverse harmonic oscillator length 
$a_{\perp}=\sqrt{\hbar/m\omega_{\perp}}$ ($m$ is the atomic mass and $\omega_{\perp}$ is the transverse confining frequency), 
the inverse frequency $\omega_{\perp}^{-1}$, and the inverse length $(2|a_{0}|)^{-1}$, where $a_0$ is the (negative) scattering length 
of the corresponding collisionally homogeneous system. In this work, we assume a collisionally inhomogeneous condensate, such that 
the nonlinear coefficient $g$ in Eq. (\ref{1dGPE}) has a spatial dependence of the form, 
\begin{equation}  
g(z)= 1-\epsilon z^2, 
\label{gofz}
\end{equation}
in the region $z \in \{-\epsilon^{-1/2}, \epsilon^{-1/2}\}$, with $\epsilon$ being a small parameter. 
Such a setting may be realized in a lithium condensate, by applying an external 
magnetic field with a dc value corresponding to the minimum of the scattering length of 
$^{7}$Li \cite{brightexp1}, and a linear gradient, which is controlled by the parameter 
$\epsilon$ (see \cite{g2} for a detailed discussion). Note that the number of atoms $\cal{N}$ in the condensate is given by 
$\cal{N}$$=(a_{\perp}/2|a_{0}|)N$, where $N = \int_{-\infty}^{+\infty} |\phi|^2 dz$ is the integral of motion (norm) for the 
normalized GP Eq. (\ref{1dGPE}). 

As is well known, in the homogeneous limit of $\epsilon=0$, the GP equation becomes the completely integrable nonlinear Schr\"{o}dinger (NLS) 
equation, possessing bright soliton solutions. In the inhomogeneous case $\epsilon \ne 0$, 
soliton solutions can still be found and studied analytically (for sufficiently small $\epsilon$) 
by means of the adiabatic perturbation theory for solitons \cite{kima} (see below). Importantly, 
in this adiabatic regime, bright solitons feel an effective trapping potential induced by the 
inhomogeneity, in which both the center and the amplitude of the solitons perform oscillations 
when displaced from $z=0$ \cite{g2,g1}. 

The purpose of this work is to demonstrate that these persistent oscillations are associated to the 
existence of a discrete eigenvalue in the corresponding linear eigenvalue problem. This eigenvalue, 
as well as the associated eigenmode, which is actually the {\it translation} mode of the solitary wave, 
manifests itself due to the inhomogeneity-induced perturbation (see also \cite{im} for a relevant discussion concerning the so-called 
{\it internal modes} of the solitary waves in nearly-integrable systems). 
We will treat the eigenvalue problem 
analytically using an approach based on 
the general theory of perturbed Hamitlonian dynamical systems 
of the nonlinear Schr{\"o}dinger type as developed in 
\cite{kap2,pan} (see also references therein).
This way, we will derive the above mentioned discrete eigenfrequency associated with the translation mode. 
Moreover, we will demonstrate that this eigenfrequency is, 
in fact, the strength of the effective 
harmonic trap induced by the inhomogeneity, and, as such, it can directly be derived 
employing the adiabatic perturbation theory for solitons. Numerical results are found to be 
in excellent agreement with the analytical predictions.

Finally, we also study a modified 1D GP model, which takes into account the effect of dimensionality. 
In fact, we explore the so-called nonpolynomial Schr\"{o}dinger equation (NPSE) \cite{sal1}, 
which can effectively describe the longitudinal wavefunction of a truly three-dimensional 
(3D) ``cigar-shaped'' BEC. The NPSE model can be expressed in the following dimensionless form:
\begin{equation}  
i \partial_{t} \phi =\left[-\frac{1}{2}\partial_{z}^{2} - 
\frac{3g(z)|\phi|^2-2}{2(1-g(z)|\phi|^2)^{1/2}}\right]\phi, 
\label{1dNPSE}
\end{equation}
in which space, time and density are normalized as in the 1D GP Eq. (\ref{1dGPE}), 
the number of atoms is again $\cal{N}$$=(a_{\perp}/2|a_{0}|)N$,  
while $g(z)$ is given by Eq. (\ref{gofz}). We consider the 
linear eigenvalue problem for the NPSE model as well, and show that the deviation from the purely 1D regime 
results in an eigenfrequency upshift. The value of the pertinent eigenfrequency will be compared to 
the oscillation frequency of a bright soliton in the full 3D GP model. 
It is shown that the agreement between the two 
is fairly good for sufficiently small values of the normalized number of atoms $N$, 
i.e., sufficiently below the collapse threshold.

The paper is organized as follows: in section II we present the 
analytical and numerical results pertaining to the 1D GP equation. 
Then, in section III we study the effect 
of dimensionality on the translation mode's frequency. 
Finally, in section IV we summarize our findings.

\section{One-dimensional condensates}

Let us follow the approach, based on 
general theory of perturbed Hamiltonian eigenvalue problems, of 
\cite{kap2,pan} to show the existense of the translation 
mode of the collisionally-inhomogeneous 
matter-wave soliton, and calculate its eigenfrequency. First we note that in the unperturbed case of 
$\epsilon=0$, the GP Eq. (\ref{1dGPE}) possesses an exact stable 
stationary bright soliton solution of the form,
\begin{equation}
\phi_{\rm bs}(z;t)= \eta {\rm sech}[\eta (z-z_0)] \exp(-i \mu t), 
\label{brightsol}
\end{equation}
where $\eta$ is the soliton's amplitude and inverse width, $z_0$ is the soliton center, and 
$\mu=-(1/2)\eta^2$ is the soliton's chemical potential. In the case $\epsilon \ne 0$, the integrability 
is broken but Eq. (\ref{1dGPE}) is still a Hamiltonian system, with Hamiltonian 
$H = H_0 + \epsilon H_1$, where $H_0$ and $H_1$ are given by [for $g(z)$ given by Eq. (\ref{gofz})]:
\begin{equation}
H_0=\int_{-\infty}^{\infty}\frac{1}{2}(|\partial_{z} \phi|^2 - |\phi|^4)dz, \,\,\,\, 
H_1= \int_{-\infty}^{\infty}\frac{1}{2} z^2 |\phi|^4 dz.
\label{h1h2}
\end{equation}
The condition for the solution of Eq. (\ref{brightsol}) to be
sustained under the considered perturbation is that the solution
remains an extremum of the perturbed energy $H_1$ \cite{kap2}, which
in our case happens for $z_0=0$. 
The stability of the perturbed solitary wave is then determined by the location of the eigenvalues 
associated with the translation and phase invariance
(the symmetries of the unperturbed problem which control the near-zero
eigenvalues of the linearized equations). 
Having in mind that the four relevant 
eigenfrequencies were located at the origin 
$\omega=0$ of the spectral plane of eigenfrequencies $\omega \equiv \omega_r + i \omega_i$ 
(in the unperturbed system), one may follow \cite{kap2,pan} and find the new location of the 
eigenfrequencies (in the perturbed system) by means of the equation:
\begin{equation}
\det(\epsilon {\bf M} -\omega^2 {\bf D}) =0,  
\label{lambda}
\end{equation}
where the matrices ${\bf M}$ and ${\bf D}$ are given by
\begin{equation}
{\bf M} = \left(
\begin{array}{cc}
\frac{\partial}{\partial z_{0}} \left< \frac{\delta H_{1}}{\delta \phi^{\star}}, \frac{\partial \phi_{\rm bs}}{\partial z_{0}} \right> & 0 \\
0 & 0 
\end{array}
\right), 
\label{M}
\end{equation}
and 
\begin{equation}
{\bf D} = \left(
\begin{array}{cc}
\left< \frac{\partial \phi_{\rm bs}}{\partial z}, -z\phi_{\rm bs} \right> & 0 \\
0 & -\left< \phi_{\rm bs}, 2\frac{\partial \phi_{\rm bs}}{\partial \eta} \right>
\end{array}
\right). 
\label{D}
\end{equation}
In the above equations, star denotes complex conjugate, $< , >$ is the inner product, and $\delta H_{1}/\delta \phi^{\star}$ 
is the functional (Fr\'{e}chet) derivative. Note that the matrix ${\bf M}$ is generally diagonal, with nonzero elements $m_{11}$ 
and $m_{22}$ acounting, respectively, for 
the perturbation-induced breaking of the translational and phase invariance. However, in our case, the considered form 
of the inhomogeneous nonlinearity does not break the phase invariance of the system and, as a result, $m_{22}=0$. 
As we will show below, a consequence of the phase invariance of the system is the appearance of a double zero eigenfrequency in the 
linear spectrum. 

The nonzero elements of the matrices ${\bf M}$, ${\bf D}$ can be directly calculated and the results are 
$m_{11}=(2/3)\eta^3$, $d_{11}=\eta$ and $d_{22}=-\eta^{-1}$. Thus, Eq. (\ref{lambda}) leads to the following simple algebraic equation,
\begin{equation}
\omega^{2}\left( \frac{2}{3}\epsilon \eta^{2}-\omega^{2} \right)=0. 
\label{ef}
\end{equation}
Equation (\ref{ef}) provides the double zero eigenfrequency $\omega^2=0$ reflecting the phase invariance of the system, 
as well as the new location of the eigenfrequencies which were associated to the translational invariance (which is now broken due to 
the presence of the spatially 
inhomogeneous nonlinearity). In fact, the latter pair of eigenfrequencies provides the 
frequency of the translation mode (which will be called $\Omega$) which is given by $\Omega^{2} = (2/3)\epsilon \eta^{2}$. 
To this end, using the relation $\mu=-(1/2)\eta^2$, the latter result can be expressed as 
\begin{equation}
\Omega= \sqrt{-\frac{4}{3} \epsilon \mu}.
\label{w_panos}
\end{equation}

It is interesting to note that an alternative analytical approach can be used to obtain the value of 
the eigenfrequency. This approach is based on the adiabatic perturbation theory for solitons \cite{kima}, 
which states that approximate soliton solutions, characterized by parameters that are unknown functions of time, 
can still be found for the perturbed system (\ref{1dGPE}). Following the methodology 
expounded in Refs. \cite{g1,fka,g2}, it is straightforward to find that 
the soliton's center evolves according to the following equation of motion,
\begin{equation} 
\frac{d^2 z_{0}}{dt^2}
=-\frac{2}{3}  \epsilon \eta^{2}(0) z_{0} 
\label{em}
\end{equation}
where $\eta(0)$ is the initial soliton amplitude. Taking into regard that $\mu=-(1/2)\eta^{2}(0)$, 
it is readily found that Eq. (\ref{em}) becomes $d^{2}z_{0}/dt^{2}=-\Omega^2 z_{0}$, where 
$\Omega$ is given by Eq. (\ref{w_panos}). This means that the soliton center 
behaves like a Newtonian unit-mass particle in the presence of the effective trapping potential 
$V_{\rm eff} = (1/2) \Omega^2 z_{0}^2$. Thus, when displaced from the trap's center ($z_{0}=0$), the soliton 
will perform harmonic oscillations with frequency $\Omega$, which is nothing but the eigenfrequency 
of the translation mode of the solitary wave. This result indicates the physical significance of the 
translation mode's eigenfrequency, which is the same as the soliton oscillation frequency in the effective 
trapping potential induced by the inhomogeneous interactions. 

The above analytical predictions have been checked by two different types of numerical simulations, in which 
$\Omega$ was directly derived by a linear stability analysis of Eq. (\ref{1dGPE}), or 
obtained as the oscillation frequency of the bright soliton in the framework of the 
GP model of Eq. (\ref{1dGPE}). 

\begin{figure}[h]
\begin{center}
\includegraphics[width=7.3cm,height=6cm]{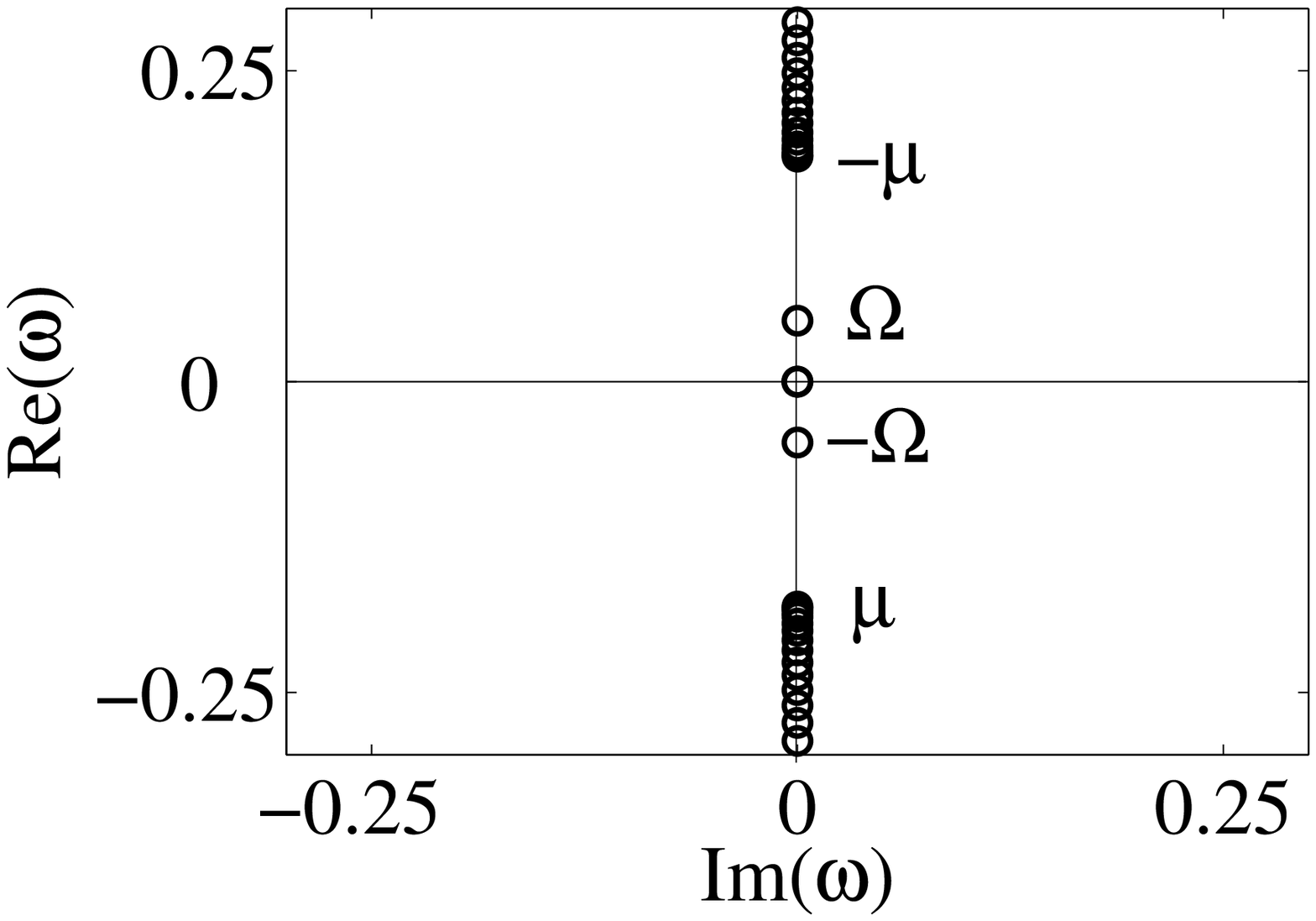}
\includegraphics[width=7cm,height=6cm]{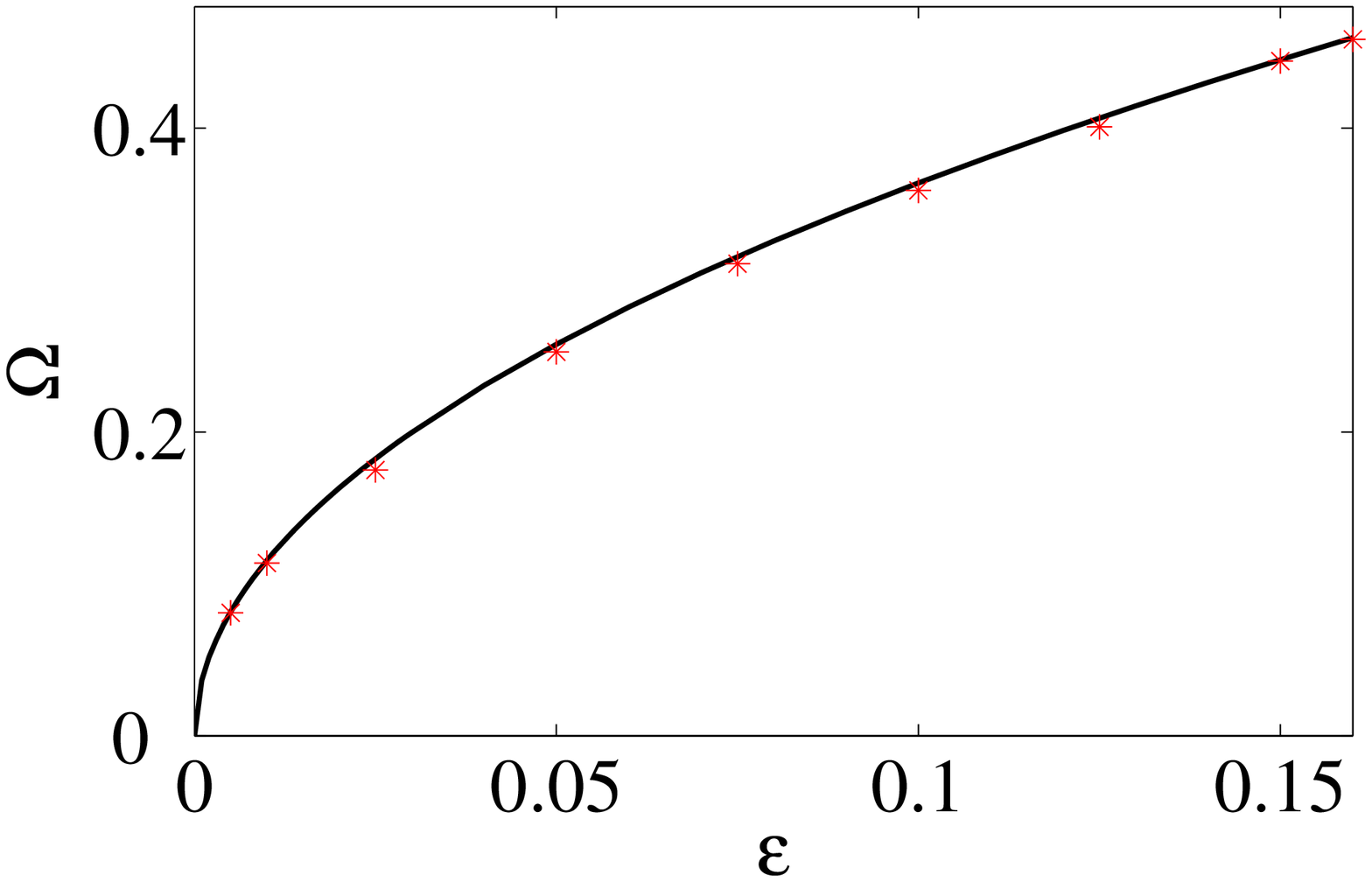}
\end{center}
\caption{(Color online) Left panel: The linear spectrum of a bright soliton of the GP Eq. (\ref{1dGPE}) 
for $\epsilon=0.01$, $N=1.2$, and $\mu=-0.18$. The bright soliton's translation mode has 
an eigenfrequency $|\Omega|=0.049$ in excellent agreement with the prediction of Eq. (\ref{w_panos}).
Right panel: Solid line shows the translation mode's frequency as a function of 
$\epsilon$ obtained by the BdG Eqs. (\ref{BdG1_GPE})-(\ref{BdG2_GPE}), 
as well as the theoretical prediction of Eq. (\ref{w_panos}) for $N=2.8$ and $\mu=-1$; the two results 
are identical and the respective curves cannot be distinguished from each
other. Stars depict 
the soliton oscillation frequency obtained by the numerical integration of Eq. (\ref{1dGPE}). 
}
\label{fig1}
\end{figure}

Let us first discuss the results obtained by the linear stability analysis, which can be performed 
upon considering small perturbations around the unperturbed soliton of the form 
\begin{equation}
\phi(z,t)=\left[\phi_{\rm bs}(z)+\epsilon\left(u(z)e^{-i\omega t}+
\upsilon^{\ast}(z)e^{i\omega^{\star} t}\right)\right]e^{-i\mu t},
\label{ansatz}
\end{equation}
where $u$ and $v$ represent the normal modes oscillating at eigenfrequencies 
$\pm \omega$. Substituting Eq. (\ref{ansatz}) into 
Eq. (\ref{1dGPE}), we obtain the following {\it Bogoliubov-de Gennes} (BdG) equations  
(valid to leading order in the small parameter $\epsilon$):
\begin{eqnarray}
\omega u &=& \left[-\frac{1}{2} \partial_{z}^{2} - \mu - F_{1}(\phi_{\rm bs}^{2}) \right] u - 
F_{2}(\phi_{\rm bs}^{2})\upsilon, 
\label{BdG1_GPE} \\
-\omega \upsilon &=& \left[-\frac{1}{2}\partial_{z}^{2} - \mu - F_{1}(\phi_{\rm bs}^{2}) \right] \upsilon - 
F_{2}(\phi_{\rm bs}^{2})u,  
\label{BdG2_GPE}
\end{eqnarray}
where $F_{1}(\phi_{\rm bs}^2)=2g(z)\phi_{\rm bs}^2$ and 
$F_{2}(\phi_{\rm bs}^2)=g(z)\phi_{\rm bs}^2$, for our real solitary wave
solutions $\phi_{\rm bs}$.
The above BdG equations have been solved numerically to find the eigenfrequencies, and the resulting spectral 
plane ${\rm Re}(\omega)$--${\rm Im}(\omega)$ is shown in the left panel of Fig. \ref{fig1} 
for $\epsilon=0.01$, $N=1.2$, and $\mu=-0.18$. 
Note that the eigenfrequencies appear in pairs due to the Hamiltonian 
nature of the system under consideration. As shown in the left panel of Fig. \ref{fig1} 
there exists a pair of eigenfrequencies at the origin (corresponding to the 
symmetry associated with the phase invariance), 
as well as a pair of  eigenfrequencies located at $\pm \Omega$, i.e., in the gap between zero and the continuous branch $|\omega|>-\mu$. 
The latter pair is the translation mode's eigenfrequency appearing due to the presence of the 
inhomogeneous interactions that break the translational 
invariance of the system. 

Moreover, in the right panel of Fig. \ref{fig1} (see solid line) we plot $\Omega$ as a function of $\epsilon$, for $N=2.8$ and $\mu=-1$. 
It is clear that the eigenfrequency $\Omega$ obtained by solving the BdG equations 
perfectly follows the square-root law of Eq. (\ref{w_panos}) (as well as the prediction of the perturbation theory for solitons);  
in fact, the respective curves are identical and cannot be distinguished from 
each other. Note that in this figure $\epsilon \in (0,0.16)$ 
so that the spatially inhomogeneous nonlinearity remains attractive (see also the discussion below).

\begin{figure}[tbp]
\begin{center}
\includegraphics[width=7.3cm,height=6.3cm]{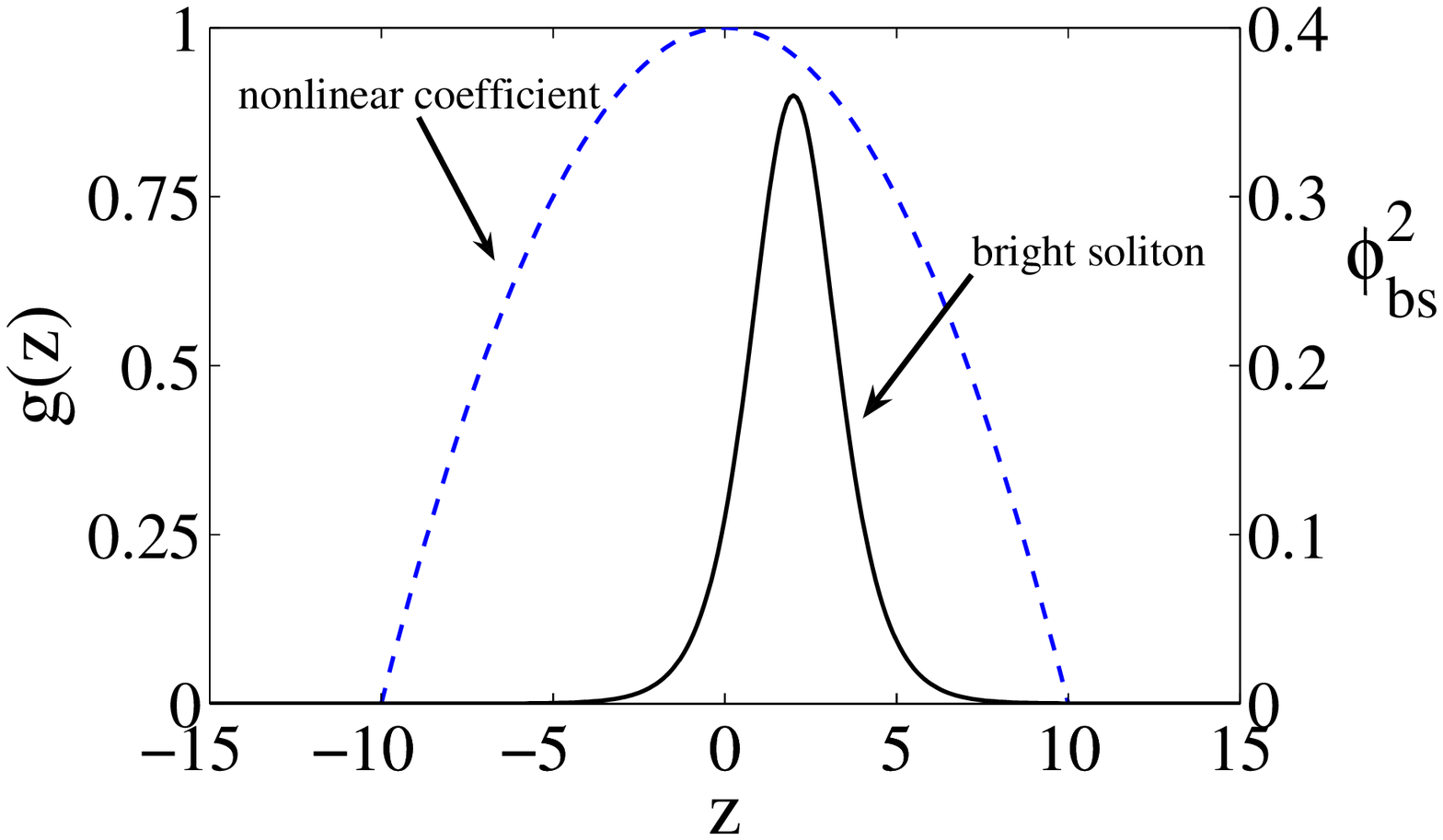}
\end{center}
\caption{(Color online) Solid line shows the density of a bright matter-wave soliton with 
$N=1.2$ and $\mu=-0.18$ (the parameter values are the ones used in Fig. \ref{fig1}); 
the soliton is initially placed at $z_{0}=2$. Dashed line shows the nonlinear coefficient $g(z)$ for $\epsilon=0.01$. 
The initial displacement of the soliton, $z_0$, which sets the amplitude of the soliton oscillation, is such that the 
inhomogeneous nonlinearity remains attractive in the region where the soliton oscillation takes place. 
}
\label{fign}
\end{figure}

Next, we have numerically integrated the GP Eq. (\ref{1dGPE}) with the initial condition found using a fixed point algorithm 
(Newton-Raphson), with the initial guess being the soliton solution in Eq. (\ref{brightsol}) at $t=0$. This ``exact'' stationary 
soliton solution was subsequently displaced so that $z_{0} \ne 0$. Note that for a fixed value of the chemical potential 
$\mu$ (or soliton width $\eta^{-1}$), the displacement of the soliton is such that the soliton oscillates in the region where the 
nonlinearity is attractive, i.e., in the interval $z \in \{-\epsilon^{-1/2}, \epsilon^{-1/2}\}$. An example of the initialization of the system 
is shown in Fig. \ref{fign}, where both the inhomogeneous nonlinear coefficient $g(z)$ and the density of a bright matter-wave 
soliton (displaced from the effective trap center) are shown. 
The resulting frequencies of the soliton oscillations are depicted by stars in the right panel of Fig. \ref{fig1}, 
which are clearly located very close to the solid line $\Omega(\epsilon)$. We conclude that there is a remarkable 
agreement between the solution of the BdG equations, the integration of the GP equation and 
the predictions of the two different perturbative approaches. 

As far as the validity of our predictions is concerned, we note the following: 
if the soliton width, $\eta^{-1}$, is sufficiently smaller than the characteristic width of the inhomogeneity, $\epsilon^{-1/2}$, 
the soliton satisfies the relevant predictions very accurately in the 
small and intermediate oscillation amplitude regime. 
This is expected to occur due to the robustness of the soliton, 
captured by the 
perturbation theory for solitons, in the perturbed--inhomogeneous--system, 
and the validity of the (linear) BdG analysis for small-amplitude oscillations.
However, in the case $\eta^{-1} \sim \epsilon^{-1/2}$, or/and for large 
amplitude oscillations, the soliton evolves under 
a strong inhomogeneous perturbation and, as a result, nonlinear effects, as 
well as emission of radiation, become 
important (see, e.g., Fig. 2 in \cite{g2}). In such cases, our assumptions 
cease to be valid and, as a result, 
our analytical (perturbative) approaches should not be
expected to agree with the numerical results.


\section{Cigar-shaped condensates}

In many experimentally relevant situations, the transverse confinement of the condensates is not 
sufficiently tight and, as a result, deviations from 1D are quite relevant. In such cases, the 
condensates are ``cigar-shaped'' and their mean-field description requires the consideration of either the 
3D GP equation, or other effectively 1D models \cite{sal1,kam,boris}. Here, we consider the 
NPSE model of Eq. (\ref{1dNPSE}), which has successfully been used to describe recent experimental 
results \cite{markus1}.

\begin{figure}[tbp]
\includegraphics[width=7.3cm,height=6cm]{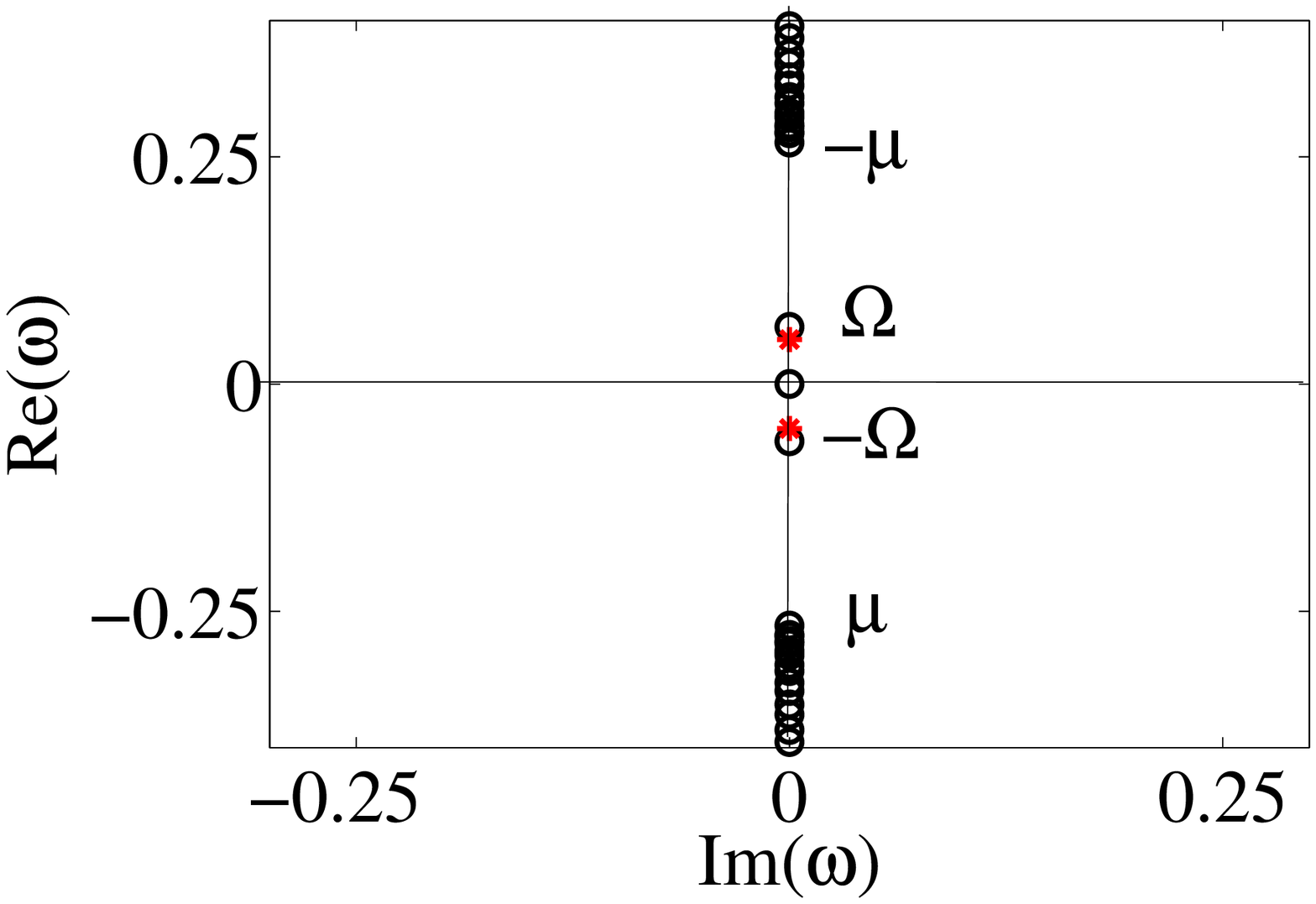}
\includegraphics[width=7cm,height=6cm]{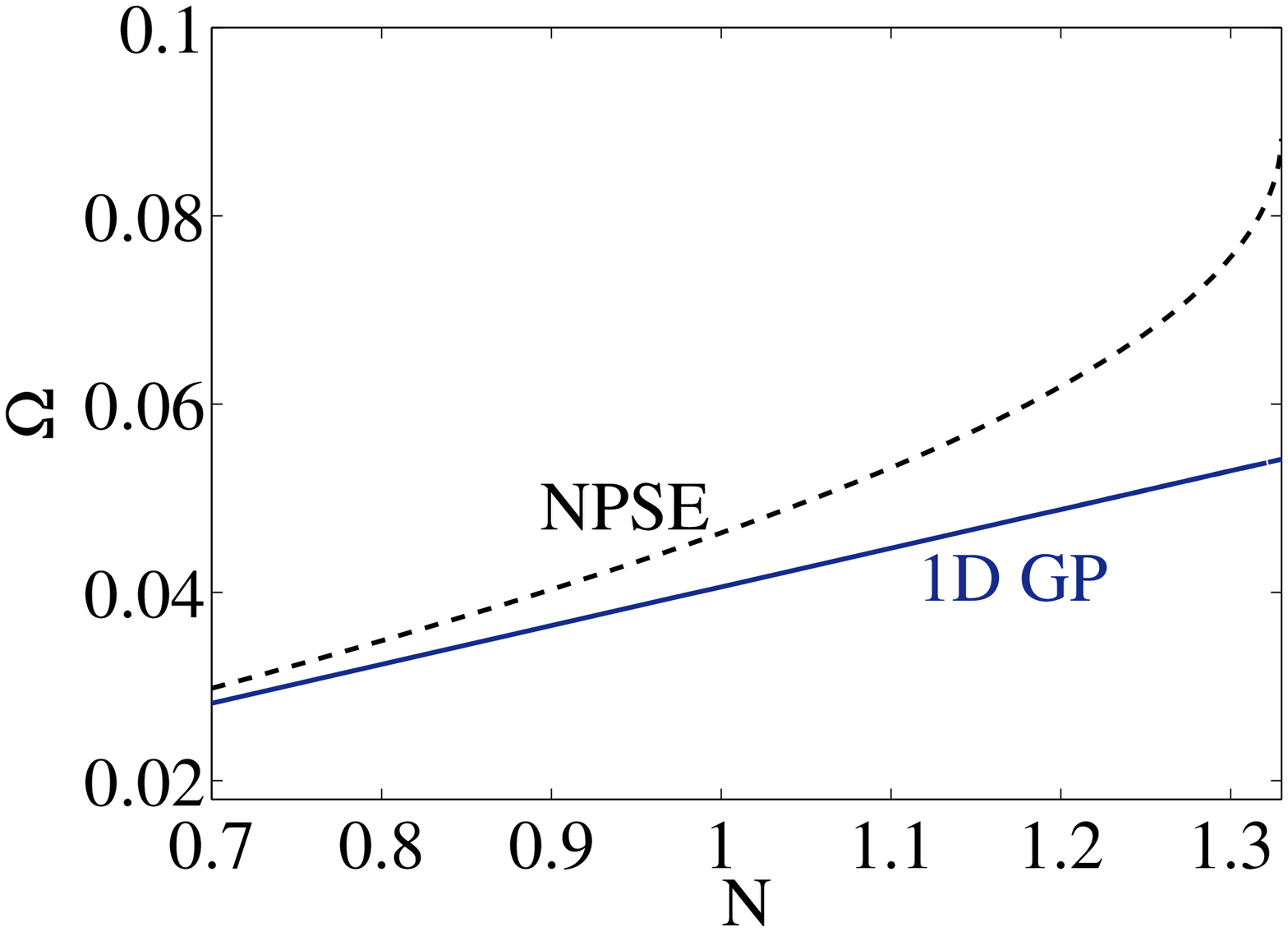}
\caption{(Color online) Left panel: Similar to the left panel of Fig. \ref{fig1}, 
but for the NPSE model of Eq. (\ref{1dNPSE}). The inhomogeneity parameter is the same 
($\epsilon=0.01$) and the chemical potential is chosen ($\mu=0.762$) 
so that the norm of the bright soliton be the same in both GP and NPSE models. 
The translation mode's frequency is up-shifted, i.e., $|\Omega|=0.062$, as compared to 
the respective value calculated with 1D GP equation (depicted by stars).
Right panel: The translation mode's frequency $\Omega$ as a function of the norm $N$, 
calculated by the BdG analysis of the NPSE model (dashed line) and the 1D GP model (solid line).  
The eigenfrequency $\Omega$ is clearly upshifted due to increase of the dimensionality. 
The inhomogeneity parameter is again $\epsilon=0.01$.
}
\label{fig2}
\end{figure}

Following the same procedure as in the case of the GP Eq. (\ref{1dGPE}), 
we introduce the ansatz of Eq. (\ref{ansatz}) in Eq. (\ref{1dNPSE}) to obtain 
BdG equations similar to Eqs. (\ref{BdG1_GPE})-(\ref{BdG2_GPE}), 
but with the functions $F_1$ and $F_2$ given by:
\begin{eqnarray}
F_{1}(\phi_{\rm bs}^{2}) &=& \frac{9g^{2}(z)\phi_{\rm{\rm bs}}^{4}
-14g(z)\phi_{\rm{\rm bs}}^{2}+4}{4[1-g(z)\phi_{\rm bs}^2]^{3/2}},
\label{F1} \\
F_{2}(\phi_{\rm bs}^{2}) &=& \frac{3g^{2}(z)\phi_{\rm{\rm bs}}^{4}
-4g(z)\phi_{\rm{\rm bs}}^{2}}{4[1-g(z)\phi_{\rm bs}^2]^{3/2}}.
\label{F2}
\end{eqnarray}
The eigenfrequency spectrum of the NPSE model has a form similar to the one pertaining to the 
GP model. However, as shown in the left panel of Fig. \ref{fig2}, the translation mode's eigenfrequency is clearly
upshifted: For the same value of the inhomogeneity parameter ($\epsilon=0.01$), and the same 
norm ($N=1.2$), we find that $\Omega=0.062$, which is $\approx 27 \%$ upshifted as compared to 
the corresponding value obtained for the GPE model ($\Omega=0.049$). 

We have found that this effect, i.e., the upshift of the translation mode's eigenfrequency 
due to the increase of dimensionality, is a generic feature of the system, as shown in 
the right panel of Fig. \ref{fig2}. 
In this figure, the eigenfrequency $\Omega$ is given as a function of the norm $N$ (for $\epsilon=0.01$) 
for both the NPSE model (dashed line) and the 1D GP model (solid line); the 
latter is apparently a near-straight line [due to the validity of Eq. (\ref{em}) and the fact that $N=2\eta$ 
for the bright soliton]. The range of values 
of $N$ is such that $N \le 1.33$, i.e., below the collapse threshold $N_{\rm c} = 1.33$ \cite{sal2}, and 
$N \ge 0.7$, so that the inhomogeneous nonlinearity remains attractive in the region where the soliton oscillation takes place. 
The latter requirement is due to the fact that $N$ is inversely proportional to the soliton width (see also Fig. \ref{fign}).

\begin{figure}[tbp]
\includegraphics[width=7cm,height=6cm]{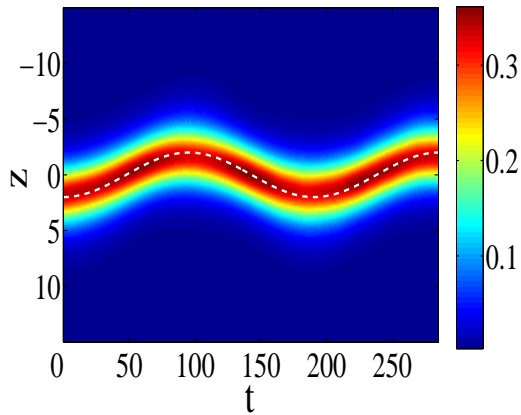}
\caption{(Color online) Spatiotemporal contour plot of the soliton's longitudinal density for $N=0.757$, 
$\mu=0.92$, and initial soliton position $z_{0}=2$. The dashed line shows the harmonic oscillation 
of the soliton center, with a frequency predicted by the BdG analysis of the NPSE model. Note that for the 
chosen value of $\epsilon$ the inhomogeneous nonlinearity remains attractive for $-10<x<10$.
} 
\label{fig3}
\end{figure}

To corroborate the validity of the NPSE analysis for the higher dimensional
case, we have numerically integrated the pertinent 3D GP equation
\begin{equation}
i \partial_{t} \psi =\left[-\frac{1}{2} \nabla^{2} + \frac{1}{2}r^2 - g(z)|\psi|^2 \right]\psi, 
\label{3D_GPE}
\end{equation}
in which $\nabla^{2} \equiv r^{-1} \partial_{r}(r \partial_{r})+ \partial_{z}^{2}$, and the 
additional trapping potential term $(1/2)r^{2} \psi$ in the transverse direction has been incorporated. 
The initial condition was obtained by a relaxation method, using the initial guess 
\begin{equation}
\psi(t=0)= \frac{\sqrt{2}}{\sigma}\exp\left(-\frac{r^2}{2\sigma^2} \right) \phi_{\rm bs}(z)
\label{ic}
\end{equation}
where the width of the Gaussian in the transverse direction is $\sigma=\sqrt{1+|\phi_{\rm bs}|^2}$  
(as per the ansatz used in \cite{sal1}) and $\phi_{\rm bs}$ is the soliton profile of Eq. (\ref{brightsol}). 
An example of the simulations is shown in Fig. \ref{fig3}, where the spatio-temporal plot of the 
soliton's longitudinal density is shown (for $N=0.757$ and $\mu=0.92$) and compared to the prediction of 
the BdG analysis of the NPSE model (dashed line). It can be observed that 
the agreement between the two is very good: 
The BdG analysis predicts an translation mode of frequency $\Omega=0.033$, while the result of the 3D simulation 
shows that the oscillation frequency is $0.03307$, with the error being $0.2\%$). We note in passing that in the purely 
1D regime, the BdG analysis of the 1D GP model predicts an eigenfrequency $\Omega=0.02989$, or 
$10\%$ discrepancy from the 3D result in this case. We should also remark 
that the deviation of the 1D NPSE result for the 
oscillation frequency from the fully 3D GP equation becomes worse
for larger values of the norm $N$, i.e., 
when approaching the collapse threshold of $N=1.33$. 
For example, a discrepancy of $8\%$ was found for 
$N=1$ (the BdG analysis of the NPSE predicted $\Omega=0.0457$ while the 3D GP model provided an oscillation 
frequency $0.0498$).

\section{Brief Summary}

We have investigated the static and dynamic properties 
of matter-wave bright solitons in a parabolic collisionally inhomogeneous
environment. It is shown that a translation mode of 
the soliton can be excited in this setting and we have found analytically its frequency. 
Numerical results, based on the relevant Bogoliubov-de Gennes equations, were found to be in 
excellent agreement with the analytical predictions. 
Moreover, we have shown that in the purely 1D 
setting, the oscillation frequency obtained in the framework of the mean-field Gross-Pitaevskii model coincides 
with the translation mode's frequency. Deviations from 1D were also considered upon analyzing the 
nonpolynomial Schr\"{o}dinger model and it was found that the translation mode's frequency is upshifted.
Numerical results obtained by the 3D Gross-Pitaevskii equation demonstrated that the oscillation 
frequency of the bright soliton in this ``cigar-shaped'' setting is generally underestimated by the 
nonpolynomial Schr\"{o}dinger model, which, in turn, is underestimated
by the 1D Gross-Pitaevskii equation for the same total number of particles. 
However, for numbers of atoms sufficiently below 
the collapse threshold the BdG analysis of the nonpolynomial 
Schr\"{o}dinger model accurately predicts 
the soliton's oscillation frequency in the 3D setting.

This work has been partially supported from ``A.S. Onasis'' Public Benefit Foundation (G.T.) and 
the Special Research Account of the University of Athens (P.N., G.T. and D.J.F.).

\end{document}